\documentclass[twocolumn, prc, amssymb, superscriptaddress,aps,preprintnumbers,amsmath,floatfix]{revtex4}
\usepackage{multirow}
\usepackage{isotope}
\usepackage{graphicx}
\usepackage{color}

\newcommand{\ket}[1]{\left| #1 \right\rangle}

\newcommand{\rket}[1]{\left\| #1 \right\rangle}
\newcommand{\rbra}[1]{\left\langle #1 \right\|}

\begin{document}

\title{The electric quadrupole channel of the 7.8 eV $\isotope[229]{Th}$ transition }


\author{Pavlo V. \surname{Bilous}}
\email{Pavlo.Bilous@mpi-hd.mpg.de} \affiliation{Max-Planck-Institut f\"ur
Kernphysik, Saupfercheckweg 1, D-69117 Heidelberg, Germany}

\author{Nikolay \surname{Minkov}}
\email{nminkov@inrne.bas.bg}
\affiliation{Institute of Nuclear Research and
Nuclear Energy, Bulgarian Academy of Sciences, Tzarigrad Road 72, BG-1784
Sofia, Bulgaria}

\author{Adriana P\'alffy}
\email{Palffy@mpi-hd.mpg.de} \affiliation{Max-Planck-Institut f\"ur
Kernphysik, Saupfercheckweg 1, D-69117 Heidelberg, Germany}


\date{\today}

\begin{abstract}
The unique isomeric transition at 7.8 eV in $\isotope[229]{Th}$ has a magnetic dipole ($M1$) and 
an electric quadrupole ($E2$) multipole mixing. So far, the $E2$ component has been widely disregarded. Here,
we investigate the nuclear physics nature and the impact of the $E2$ decay channel for the nuclear coupling to the atomic shell based on the newest theoretical predictions for 
the corresponding reduced nuclear transition probabilities.
Our results show that the contribution of the $E2$ channel is dominant or at least of the same order of magnitude for internal conversion or electronic bridge transitions 
involving the atomic orbitals $7p$, $6d$ and $5f$. Notable exceptions are the internal conversion of the $7s$ electron and the electronic bridge 
 between the electronic states $7s$ and $7p$, for which the $M1$ component dominates by two to three orders of magnitude. Caution is therefore advised when considering 
 isomeric excitation or decay via nuclear coupling to the atomic shell, as the involved orbitals determine which multipole transition component dominates.

\end{abstract}

\maketitle

\section{Introduction}
The nuclear transition with the lowest energy known at present connects the 7.8 eV isomer in
$\isotope[229]{Th}$ to the nuclear ground state. The first hints on the existence of the isomeric state
date back to 1976 when a ground state  doublet
was proposed to explain the observed decay cascades in gamma-ray spectroscopy
experiments in $^{229}$Th~\cite{KrogerReich1976}.

The nuclear structure of $^{229}$Th is  representative for the heavy
nuclei in the actinide region where pronounced collectivity with possible presence of
octupole (reflection-asymmetric) deformation is typically observed~\cite{BN96}. A recent
theoretical study suggests that the observed nuclear structure complexity in this isotope  is determined by the
combined effects of the collective quadrupole-octupole vibration-rotation motion of the nucleus,
the single-particle (s.p.) motion of the odd, unpaired nucleon and the Coriolis interaction
between the latter and the nuclear core~\cite{Minkov_Palffy_PRL_2017}. Within this concept
it was suggested that the appearance of the $3/2^{+}$ isomeric state at
approximately $7.8$ eV as an almost degenerate counterpart of the $5/2^{+}$ ground state can only be accounted for
by an extremely fine interplay between all involved collective and s.p.
degrees of freedom. 

The isomeric transition energy $E_\mathrm{m}=7.8$ eV, corresponding to a wavelength of
approximately~160 nm, lies in the range of vacuum ultraviolet (VUV) lasers. This renders
possible a number of exceptional applications  such as a nuclear optical frequency
standard~\cite{Peik_Clock_2003,Campbell_Clock_2012,Peik_Clock_2015}, nuclear
laser~\cite{Tkalya_NuclLaser_2011} or coherent control of the nuclear
excitation~\cite{Liao_Coherent_2012,Das2013}. From the experimental point of view, all
these applications require a more  precise determination of the transition energy, since 
the presently used value of  $E_\mathrm{m}=7.8\pm 0.5$ eV (i) stems from an indirect measurement 
performed in 2007 via energy-resolved detection of x-rays in the $\isotope[229]{Th}$ decay cascade~\cite{Beck_78eV_2007,Beck_78eV_2007_corrected}, (ii) is
at present questioned due to the use of uncertain branching ratios in the experimental data analysis and (iii) has a rather large error bar.
In the past decade, several attempts of direct photoexcitation of the isomeric state around the 7.8 eV value at broadband light sources failed~\cite{Jeet_PRL_2015,Yamaguchi2015}. The direct evidence of the isomer was gained by observing its non-radiative decay \cite{Wense_Nature_2016} in the process of  internal
conversion (IC), i.e., the radiationless nuclear excitation energy transfer to an electron from the atomic shell followed by ionization. 
Furthermore, the lifetime of the isomeric state in neutral thorium due to the IC decay was determined to be
10~microseconds~\cite{Seiferle_PRL_2017}. 
Finally, a very recent measurement was able to identify the hyperfine structure of the isomeric state~\cite{Thielking2017}, which opens the future for clock interrogation schemes based on the electronic hyperfine structure in  $\isotope[229]{Th}$ ions.

These recent experimental findings show the important role that the atomic shell plays for the decay mechanisms of the $\isotope[229]{Th}$ isomer.
The excitation energy of 7.8$\pm 0.5$ eV is larger
than the first ionization potential of thorium at 6.3 eV, but however  smaller than that of Th$^+$
ions at approximately~12~eV~\cite{HerreraSancho_PRA_2012}. IC of the outmost electron
is, therefore, possible and many orders of magnitude more probable than the radiative decay
of the isomeric state in neutral thorium atoms. The IC coefficient, i.e., the ratio of the IC and
radiative rates is expected to be approximately $10^9$~\cite{Karpeshin2007}, such that atomic outer shell electrons play the dominant role in the 
isomer decay. Apart from IC, also the electronic bridge process (EB) may play an important
role in the excitation or decay of the isomer~\cite{Krutov_JETPLett_1990,
StrizhovTkalya_JETP_1991,Karpeshin_PRL_1999,Kalman_PRC_2001,PorsevFlambaum_Brige3+_PRA_2010,PorsevFlambaum_Brige1+_PRA_2010,PorsevFlambaum_Brige_PRL_2010,Bilous_NJP_2018}. EB is a second order process in which the nuclear excitation is transferred to the electronic shell with the additional emission or absorption of a photon.
IC from excited  electronic states in Th$^+$ and Th$^{2+}$ ions has also been recently
investigated~\cite{Bilous_PRA_2017}.

The transition between the isomeric and the ground state occurs via a magnetic dipole ($M1$)
and an electric quadrupole ($E2$) multipole mixing. The $M1$ component has been assumed
so far to play the dominant role in the isomeric transition.  The  reason for this is
two-fold. First, the VUV laser-nucleus interaction which lays at the core of the most
appealing applications involving $\isotope[229\mathrm{m}]{Th}$ would proceed via the radiative
channel. The radiative decay rates however are proportional  to $E_\mathrm{m}^3$ for dipole and
$E_\mathrm{m}^5$ for quadrupole  multipolarity, respectively. With the isomeric transition energy $E_\mathrm{m}$ being very small, the additional
multiplicative factor $E_\mathrm{m}^2$ suppresses the $E2$ component by more than ten
orders of magnitude. This does not hold however for the IC or EB rates, which do not have
an explicit dependence on the transition energy. 

Second, so far the $E2$
component was considered to be negligible even for  non-radiative processes involving the
coupling to the atomic shell. For instance, 
Refs.~\cite{Dyk98,Tkalya_PRC_2015} conclude that the ratio of the $E2$ and $M1$ IC rates
should be smaller than $10^{-3}$ for the thorium atom, presumably considering the IC of the
$7s$ electron, based on the theoretical values $B(M1)\simeq 10^{-2}$~W.u. and $B(E2)\simeq 10$~W.u.
The latter approximate theoretical values for the reduced transition probabilities were deduced using 
branching ratios (Alaga rules~\cite{Alaga55}) from
observed decays of neighboring levels. EB calculations~\cite{PorsevFlambaum_Brige3+_PRA_2010} considering several atomic orbitals conclude
that the $E2$ contribution is negligible based on the values $B(M1)=4.8\times 10^{-2}$~W.u.
and $B(E2)=1$~W.u. However,  the most recent theoretical predictions~\cite{Minkov_Palffy_PRL_2017} for the reduced
transition probabilities have provided the  values $B(M1)=0.0076$~W.u. and $B(E2)=27$
W.u., which are smaller for the $M1$ transition and larger
for the $E2$ transition, respectively, compared to the previous values used for the
conclusions quoted above. A smaller $M1$ reduced transition probability is consistent with
the recently observed isomeric state lifetime~\cite{Seiferle_PRL_2017} considering an IC
coefficient of $10^9$. It should be noted that theoretical $B(M1)$ values smaller than those used in Ref.~\cite{PorsevFlambaum_Brige3+_PRA_2010,Tkalya_PRC_2015} but still larger by a factor of two than the ones
in Ref.~\cite{Minkov_Palffy_PRL_2017} were obtained through the quasiparticle-plus-phonon
model~\cite{Gulda02,Ruch06}. 

In this paper, we question the existing paradigm and investigate the existence and the impact of the $E2$ multipole mixing for the
$^{229}$Th isomeric transition from a nuclear and atomic physics perspective. From the
nuclear physics point of view, the $E2$ component is related to the strong collectivity of the
heavy thorium nucleus, which dominates its low-lying spectrum. Closer to atomic physics, we perform IC and EB
calculations  considering the most recent state-of-the-art theoretical
predictions for the reduced transition probabilities for the isomer~\cite{Minkov_Palffy_PRL_2017}. Our results
clearly show  that the coupling of the atomic
shell to the nucleus is not always negligible and can be even dominated by the $E2$
component for specific atomic orbitals. This holds true for the $5f$ and $6d$ orbitals, which
are of relevance for IC from excited electronic state and for EB processes. We formulate valid
criteria upon which the $E2$ component needs to be considered and investigate the
consequences of the multipole mixing for the experimental effort dedicated at present to the
$\isotope[229]{Th}$ isomeric transition.

The paper is structured as follows. Sec.~\ref{nucl_hardcore} discusses the
origin of electromagnetic multipole mixing in the nuclear structure model
describing the low-lying positive- and negative-parity excited levels and
transition probabilities observed in the $\isotope[229]{Th}$  nucleus. The
theory of the nuclear transition coupling to the atomic shell in the
processes of IC and EB is briefly summarized in Sec.~\ref{atomic_coupling}.
Numerical results and discussions follow in Sec.~\ref{numres}. Concluding
remarks are given at the end of the paper. Atomic units ($\hbar=e=m_e=1$) are used
throughout the paper unless otherwise specified.


\section{Nuclear structure background of the isomeric transition \label{nucl_hardcore}}
The actinide nuclei and in particular the even-odd isotopes among them present a rich nuclear
structure. A model approach capable to incorporate the shape-dynamic properties together
with the intrinsic structure characteristics typical for the actinide nuclei has been under
development in the last decade~\cite{b2b3mod,b2b3odd,qocsmod,WM10,MDSS09,MDSS10, NM13}.  It
considers a collective quadrupole-octupole (QO) vibration-rotation motion of the nucleus
which in the particular case of odd-mass nuclei is coupled to the motion of the single (odd)
nucleon within a reflection-asymmetric deformed potential. The collective motion is described
through the so-called coherent QO mode (CQOM) giving raise to the quasi parity-doublet
structure of the spectrum~\cite{b2b3mod,b2b3odd}, whereas the single-particle (s.p.) one is
determined by deformed shell model (DSM) with reflection-asymmetric Woods-Saxon
potential~\cite{qocsmod} and pairing correlations of BCS type included as in
Ref.~\cite{WM10}. The Coriolis interaction between CQOM and the odd nucleon was
originally considered in~\cite{MDSS09,MDSS10}, whereas the effect of Coriolis decoupling
and $K$-mixing on the rotation-vibration levels, with $K$  the projection of the angular
momentum on the intrinsic nuclear symmetry axis, was taken into account in~\cite{NM13}.

All the model aspects outlined above have been assembled together in
Ref.~\cite{Minkov_Palffy_PRL_2017} in a detailed nuclear-structure-model description of
the low-lying positive- and negative-parity excited levels and transition probabilities observed
in $^{229}$Th to predict the $B(M1)$ and $B(E2)$ reduced probabilities for the radiative
decay of the 7.8 eV $K=3/2^{+}$-isomer to the $K=5/2^{+}$ ground state.  The two states are
considered as almost degenerate quasi-particle bandheads with a superposed collective QO
vibration-rotation mode giving raise to yrast $K=5/2^{+}$ and non-yrast $K=3/2^{+}$ quasi
parity-doublet structures. The isomer decay is obtained as the result of a Coriolis mixing
emerging from a remarkably fine interplay between the coherent QO motion of the core and
the single-nucleon motion within  the  reflection-asymmetric deformed
potential. Despite earlier
statements on the weakness of the Coriolis mixing~\cite{Dyk98,Tkalya_PRC_2015}, we emphasize that only because  of the
Coriolis  $K$-mixing interaction  can we  explain the presence of the {\it otherwise due to the overall axial symmetry of the problem $K$-forbidden}
 $M1$ and $E2$ transitions between the yrast and non-yrast bands.

Within this model it is also clear that the two electromagnetic multipole contributions have
different  origins. The $E2$ transition is mainly related to the collective part,
 whereas the $M1$ component emerges from the  single-nucleon
degree of freedom~\cite{Ring1980}.  Nevertheless, the collective QO mode has a
strong indirect influence on the $M1$ transition via the s.p. coupling to the nuclear core.
Vice-versa, the collective part is decisive for the $E2$ transition, however with indirect
influence from the single nucleon via the particle-core coupling. The reasoning for the
existence and the decay properties of the $^{229m}$Th state is therefore strongly related to all
nuclear structure model ingredients in Ref.~\cite{Minkov_Palffy_PRL_2017}, 
namely, the collective core, the single-nucleon motion in the deformed
potential and the Coriolis interaction. For the calculations below we will use reduced transition probability values 
$B(M1)=0.0076$~W.u. and $B(E2)=27$
W.u. predicted by this state-of-the-art model.


\section{Coupling to the atomic shell \label{atomic_coupling}}

Especially due to the very small transition energy, the atomic shell can play an important role in the decay or excitation of the isomer. The radiative channel, on the other hand, is very weak, with an expected radiative width $\Gamma_\gamma$ to transition energy ratio of approx.~$10^{-20}$. This value is calculated considering only the $M1$ channel, which is much stronger than the $E2$ one. Indeed, with the radiative rates given by~\cite{Ring1980}
\begin{equation}
\Gamma_\gamma(M1)=1.779 \cdot 10^{13} \cdot E_{\mathrm{m}}^3 \cdot B(M1)
\end{equation}
for magnetic dipole transitions and 
\begin{equation}
\Gamma_\gamma(E2)=1.223 \cdot 10^9 \cdot E_{\mathrm{m}}^5 \cdot B(E2)
\end{equation}
for electric quadrupole transitions, we obtain a ratio $\Gamma_\gamma(E2)/\Gamma_\gamma(M1)= 6.9 \cdot 10^{-10}$. In the equations above, the transition probabilities (in s$^{-1}$) are expressed as a function of $B(E2)$ in $e^2 \mathrm{fm}^4$ and $B(M1)$ in units of the nuclear magneton squared $\mu_N^2$, while the energy $E_\mathrm{m}$ is considered in MeV. The value $6.9 \cdot 10^{-10}$ is larger than the ratio of $10^{-13}$ reported in Ref.~\cite{Tkalya_PRC_2015} based on the single-particle (Weisskopf) model, but still very small. For all practical purposes, the $E2$ component can be safely neglected for the radiative decay channel.

\subsection{Internal conversion rates}

Let us now consider IC for neutral Th atoms or excited Th ions. The IC transition rate expressions can be written via the reduced nuclear transition probabilities and numerical integrals of radial electronic wave function corresponding to the electronic matrix element for the Coulomb or current-current interactions between electron and nucleus~\cite{Bilous_PRA_2017}. We present here only the final expressions, for the $M1$ contribution
\begin{eqnarray}\label{gammaicM1}
\Gamma_{\mathrm{IC}}({M1}) &=& 
\frac{8 \pi^2}{9}B(M1) \sum_{\kappa}(2j+1)\nonumber \\
&\times&
 (\kappa_i+\kappa)^2
\begin{pmatrix}
j_i & j & 1 \\
1/2 & -1/2 & 0
\end{pmatrix}^2
|R_{\varepsilon\kappa}^{\mathrm{M1}}|^2 \Lambda\, ,
\end{eqnarray}
and for the $E2$ contribution 
\begin{eqnarray}
\label{gammaicE2}
\Gamma_{\mathrm{IC}}({E2}) &=& 
\frac{8 \pi^2}{25}B(E2) \sum_{\kappa}(2j+1)\nonumber \\
&\times&
\begin{pmatrix}
j_i & j & 1 \\
1/2 & -1/2 & 0
\end{pmatrix}^2
|R_{\varepsilon\kappa}^{\mathrm{E2}}|^2 \Lambda\,
\end{eqnarray}
where $j$ ($j_i$) and $\kappa$ ($\kappa_i$) represent the total angular and the Dirac angular momentum quantum numbers for the continuum (bound) electron. The sum over the continuum partial wave Dirac quantum number $\kappa$ is performed according to the corresponding selection rules for multipolarity $\lambda L=M1$ or $E2$. The quantities $R_{\varepsilon\kappa}^{\mathrm{M1}}$ and $R_{\varepsilon\kappa}^{\mathrm{E2}}$ are the radial integrals given by
\begin{eqnarray}\label{rintegrals}
R_{\varepsilon\kappa}^{\mathrm{M1}} &=& \int_{0}^{\infty} dr \Bigl( g_{n_i \kappa_i}(r)f_{\varepsilon \kappa}(r) + g_{\varepsilon \kappa}(r) f_{n_i\kappa_i}(r) \Bigr)\, , \\
R_{\varepsilon\kappa}^{\mathrm{E2}} &=& \int_{0}^{\infty} \frac{dr}{r} \Bigl( g_{n_i \kappa_i}(r)g_{\varepsilon \kappa}(r) + f_{\varepsilon \kappa}(r) f_{n_i\kappa_i}(r) \Bigr)\, ,
\end{eqnarray}
where $g_{\beta \kappa}$ and $f_{\beta \kappa}$ are the radial wave functions of the initial (bound) and final (continuum) electron in the following notation of the total electron wave function
\begin{equation}\label{generalwf}
\ket{\beta \kappa m} =
\begin{pmatrix}
g_{\beta \kappa}(r) \Omega_{\kappa m}(\hat{r}) \\
if_{\beta \kappa}(r) \Omega_{-\kappa m}(\hat{r})
\end{pmatrix}\;,
\end{equation}
where $\Omega_{\kappa m}(\hat{r})$ are the spherical spinors and $\beta$ represents the principal quantum number $n$ for bound electron orbitals and the energy $\varepsilon$ for the continuum electronic state. The factor $\Lambda$ in Eqs.~(\ref{gammaicM1}) and (\ref{gammaicE2}) quantifies the dependence of the IC rates on the angular momentum coupling of the outer electrons and is typically of order unity. For example, considering an initial electronic configuration with three electrons, if originally the two spectator electrons have coupled angular momenta and IC expels the third electron, then $\Lambda=1$. Otherwise,  if the angular momentum of one spectator electron $j_1$ and the angular momentum of the IC electron $j_i$ are originally coupled to the angular momentum $J_0$ which is then coupled with the angular momentum of the second spectator electron $j_2$ to the total angular momentum $J_i$, we have
\begin{equation}\label{Lambda}
\Lambda = (2J_0+1)(2J_f+1)
\left\{
\begin{matrix}
j_1 & j_2 & J_f \\
j_i & J_i & J_0
\end{matrix}
\right\}^2\, ,
\end{equation}
where the curly brackets represent the $6j$-coefficient defining possible values $J_f$ of the angular momentum of the two electronic configuration after IC.  In the following we consider IC for different orbitals of a neutral Th atom, whose outer shell contains four electrons. Since our main concern is the ratio of the $M1$ and $E2$ IC rates for different specific orbitals, we will omit the factor $\Lambda$ in the following.

Apart of the respective values for the reduced transition probabilities $B(E2)$ and $B(M1)$, the important quantities entering the expressions of the IC rates are the radial integrals $R_{\varepsilon\kappa}^{\lambda\mathrm{L}}$. Their values depend on the initial bound orbital for which IC occurs and strongly influence the magnitude of the IC rate. It is also of high significance which continuous electronic states are allowed by selection rules in the sums over $\kappa$ in Eqs.~(\ref{gammaicM1}) and~(\ref{gammaicE2}).

\subsection{Electronic bridge rates}

\begin{figure*}[ht!]
\centering
\includegraphics[width=0.85\textwidth]{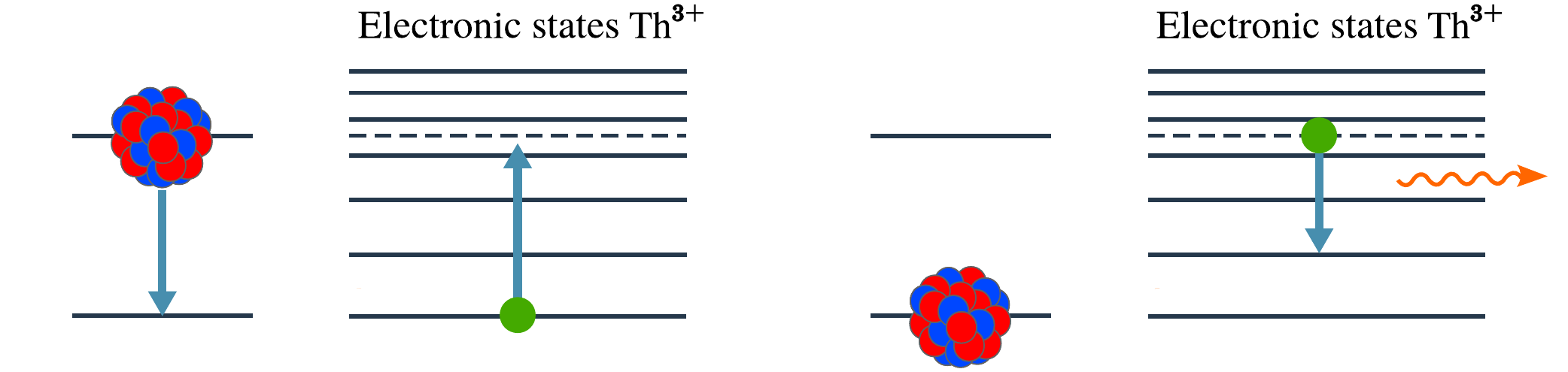}
\caption{(Color online.) A schematic illustration of electronic bridge in a monovalent ion $\isotope{Th}^{3+}$. The excited nucleus transfers its energy to a bound electron, which undergoes a transition to a virtual state (dashed line). The latter decays to a real bound state with the emission of a photon. \label{figbridge}}
\end{figure*}

EB is another process coupling the nucleus to the atomic shell~\cite{Krutov_JETPLett_1990, StrizhovTkalya_JETP_1991,Karpeshin_PRL_1999,Kalman_PRC_2001}. It occurs when the nuclear transition energy is not sufficient for IC, but is close to an atomic transition energy. In order to fulfill energy conservation, the 
transfer of the nuclear excitation to a bound electron which undergoes a transition to an excited state is accompanied by the emission or absorption of a photon. An example of EB involving the nuclear decay with an electronic transition accompanied by the emission of a photon is illustrated in Fig.~\ref{figbridge}. The EB process might play a significant role in the decay of the isomeric state in Th ions, where the IC channel is energetically closed. 
Calculations have shown that the EB process can significantly change the isomeric state lifetime if the energy $E_{\mathrm{m}}$ happens to be close to energy of $M1$ transitions of the electronic shells~\cite{PorsevFlambaum_Brige3+_PRA_2010,PorsevFlambaum_Brige1+_PRA_2010}. Less importance has been attributed so far to the $E2$ contribution~\cite{PorsevFlambaum_Brige3+_PRA_2010}. 

The ratio of the EB and radiative rates is denoted by $\beta$ and corresponds to an IC-coefficient equivalent adapted for EB. The explicit expressions are given by~\cite{PorsevFlambaum_Brige3+_PRA_2010}
\begin{eqnarray}\label{betas}
\beta(M1)=\left(\frac{\omega}{E_{\mathrm{m}}}\right)^3\frac{G^{M1}_1+G^{M1}_{12}+G^{M1}_2}{3(2J_i+1)}\;,\\
\beta(E2)=\frac{4}{k_\mathrm{m}^2}\left(\frac{\omega}{E_{\mathrm{m}}}\right)^3\frac{G^{E2}_1+G^{E2}_{12}+G^{E2}_2}{2J_i+1}\;,
\label{betas2}
\end{eqnarray}
where $\omega$ is the emitted photon frequency, $J_i$ is the initial total angular momentum of the electronic shell and the wave vector $k_\mathrm{m}$ can be expressed via the nuclear transition energy $E_\mathrm{m}$ as $k_\mathrm{m}=\alpha E_\mathrm{m}$ with the fine structure constant $\alpha$. The quantities $G_i$ with $i=1,2$ correspond to the two possible temporal sequences of virtual photon exchange and emission of a real photon, while $G_{12}$ stands for  their interference. All three terms corresponding to magnetic dipole (electric quadrupole) nuclear transition can be written with the help of the reduced matrix elements of the magnetic dipole $T^{M1}$ (electric quadrupole $T^{E2}$) operators and the electric dipole operator $D$~\cite{Johnson_book_2007} of the valence electron giving the following expressions
\begin{eqnarray}\label{g1}
G_1^{\lambda L}&=&\sum_{J_n}\frac{1}{2J_n+1}\\
 &\times& \left| \sum_{\gamma_k} 
 \frac{\rbra{\gamma_f J_f}D\rket{\gamma_k J_n} \rbra{\gamma_k J_n} T^{\lambda L} \rket{\gamma_i J_i}}
 {E_i-E_k+E_{\mathrm{m}}}
 \right|^2\nonumber
\end{eqnarray}
\begin{eqnarray}\label{g12}
G_{12}^{\lambda L}&=& 2\sum_{J_t J_n} (-1)^{J_t+J_n}
\left\{
\begin{array}{ccc}
J_i & J_t & 1 \\
J_f & J_n & L
\end{array}
\right\}
\\
 &\times & \sum_{\gamma_k} 
 \frac{\rbra{\gamma_f J_f}D\rket{\gamma_k J_n} \rbra{\gamma_k J_n} T^{\lambda L} \rket{\gamma_i J_i}}
 {E_i-E_k+E_{\mathrm{m}}}\nonumber\\
 &\times& \sum_{\gamma_s} 
 \frac{\rbra{\gamma_f J_f}T^{\lambda L}\rket{\gamma_s J_t} \rbra{\gamma_s J_t} D \rket{\gamma_i J_i}}
 {E_f-E_s-E_{\mathrm{m}}}\nonumber
\end{eqnarray}
\begin{eqnarray}\label{g2}
G_2^{\lambda L}&=&\sum_{J_n}\frac{1}{2J_n+1}\\
 &\times& \left| \sum_{\gamma_k} 
 \frac{\rbra{\gamma_f J_f}T^{\lambda L}\rket{\gamma_k J_n} \rbra{\gamma_k J_n} D \rket{\gamma_i J_i}}
 {E_f-E_k-E_{\mathrm{m}}}
 \right|^2\, .\nonumber
\end{eqnarray}
 The sums are carried out over the total angular momenta of the intermediate states $J_n$ and $J_t$ and over all other electronic quantum numbers denoted by the generic indices $\gamma_k$ and $\gamma_s$. The notation $\rbra{}\cdot\rket{}$  stands for the reduced matrix elements after application of the Wigner-Eckart theorem~\cite{Edmonds}.

\section{Numerical results \label{numres}}

In the following we present our results for IC and EB rates involving relevant orbitals of both neutral thorium and of thorium ions in several charge states. We focus our analysis on the relative contribution of the $M1$ and $E2$ multipoles. 

The calculation of the IC and EB rates requires knowledge of electronic level energies and wave functions. The spectrum of the valence electron is taken from atomic spectroscopy data~\cite{DBlevels} which provides precise values as opposed to the limited accuracy of atomic structure calculations for atoms or ions with many electrons. For evaluation of the electronic matrix elements for the IC and EB rates we use different numerical approaches. In the case of IC, the  matrix elements are evaluated using relativistic electronic wave functions which for the bound electron are obtained from a multi-configurational Dirac-Hartree-Fock (DHF) method using the GRASP2K package~\cite{grasp}. The continuum wave functions are calculated with the program \textit{xphoto} from the RATIP package~\cite{ratip}. 

EB calculations require however a different numerical approach, due to the following reason.  The expressions for the quantities $G$ in Eqs.~(\ref{g1})--(\ref{g2}) contain summations over all intermediate electronic states allowed by the corresponding selection rules, including summation over highly excited bound states and integration over continuum states. The GRASP2K package is however limited in evaluation of excited states already with principal quantum numbers approaching $n=10$ and cannot provide all needed wave functions. On the other hand, integration over continuum wave functions obtained from the RATIP package is not straightforward. We use instead the following scheme for evaluation of the required matrix elements. As a first step we calculate a few low-lying states of the valence electron using the DHF method in the frozen-core approximation. Secondly, we consider the ion to be placed into a cavity of radius $R=90\;\mathrm{a.u.}$ and build virtual orbitals with the principal quantum number up to $n=30$ via the expansion in a $B$-spline basis~\cite{Johnson_book_2007}. The result of the DHF calculation serves as in input for this second step. This procedure allows us to work with a discrete spectrum at positive energies and due to the large size of the cavity does not affect the accuracy of the calculated matrix elements considerably. While the values for the $E2$ matrix elements obtained at this stage are satisfactory, the values for the $M1$ matrix elements require further improvement~\cite{Gossel_PRA_2013,JohnsonSafronova_PRA_2017}. To achieve adequate accuracy, we take into account correlation effects between the valence electron and the frozen core using the method of random phase approximation (RPA)~\cite{Johnson_book_2007}. In contrast to the case of electric matrix elements, the obtained RPA corrections for magnetic matrix elements are often significantly larger than the corresponding DHF values. We double-checked our computation method for the $M1$ matrix elements by calculating the hyperfine structure constant $A$ for various orbitals of the Na valence electron and comparing them with existing values in literature for combined DHF and RPA calculations~\cite{Johnson_book_2007}. The results show a good agreement.

An important quantity in the calculation is the nuclear transition energy $E_{\mathrm{m}}$. We assume in the following that the isomer lies at 7.8 eV or above. A word is due here on the assumption that the nuclear transition energy is larger than 7.8 eV. A few recent results have shed doubt on the exact range of the nuclear transition energy. The presently used value $E_{\mathrm{m}}=7.8 \pm 0.5$~eV could be determined only indirectly in a calorimetric measurement by
subtraction of x-ray energy differences between neighboring nuclear levels~\cite{Beck_78eV_2007,Beck_78eV_2007_corrected}. However, the  extraction of the energy value $E_{\mathrm{m}}$ from the experimental data in Refs.~\cite{Beck_78eV_2007,Beck_78eV_2007_corrected} involves knowledge of nuclear branching ratios in the decay cascade, whose values are not so precisely known and which may add systematic shifts to the extracted isomeric energy~\cite{Tkalya_PRC_2015}. In addition, recent negative experimental results of two broadband photoexcitation attempts of the isomeric state may indicate that the transition energy lies in a different energy range~\cite{Jeet_PRL_2015,Yamaguchi2015}. Furthermore, the recent results in Ref.~\cite{Seiferle_PRL_2017} on the short lifetime of the isomer in $\isotope{Th}^+$  may be indirect evidence that the IC channel is already open and  $E_{\mathrm{m}}$ is higher than the ionization potential of $\isotope{Th}^+$, i.e. approx.~12~eV~\cite{HerreraSancho_PRA_2012}. These were the premises on which recent theoretical studies~\cite{Bilous_PRA_2017,Bilous_NJP_2018} have also considered the possibility that $E_\mathrm{m}>7.8$~eV.

\subsection{IC in the neutral Th atom}
We consider at first the isomer energy at 7.8 eV and the isomeric decay via IC in a neutral Th atom. 
 In this case, a $7s$ electron from the ground-state electronic configuration $6d^27s^2$ undergoes IC. There is also a relatively small probability that the $6d$ shell will be ionized. We calculate the IC rates for $7s$ and $6d$ electron for the ground state considering the  $M1$ and $E2$ channels. In order to compare the rates for the two multipolarities  also for the $5f$ and $7p$ outer orbitals, we consider IC not from the ground state but from the excited electronic configurations $5f6d7s^2$ at 7795 $\rm{cm}^{-1}$ (ionization of the $5f$ electron) and $6d7s^27p$ at 10783 $\rm{cm}^{-1}$ (ionization of the $7p$  electron). The corresponding $M1$ and $E2$ IC rates calculated using Eqs.~(\ref{gammaicM1}) and (\ref{gammaicE2}) are presented in  Table~\ref{NeutrAtICRates}.
%
\begin{table}[h!]
\centering
\begin{tabular}{|c|c|c|c|}
\hline
Orbital & $\Gamma_{IC}(M1)$ (s$^{-1}$) & $\Gamma_{IC}(E2)$ (s$^{-1}$) & $\Gamma_{IC}(E2)/\Gamma_{IC}(M1)$\\
\hline
$7s$ & $1.3 \cdot 10^5$ &  $3.8 \cdot 10^2$ & $ 2.9 \cdot 10^{-3} $\\
\hline
$7p_{1/2}$ & $4.2 \cdot 10^3$ & $ 5.1 \cdot 10^3$ &$ 1.2 $\\
\hline
$7p_{3/2}$ & $3.5 \cdot 10^2$ & $ 8.2 \cdot 10^3$ &$ 23 $\\
\hline
$6d_{3/2}$ & $2.3 \cdot 10^2$ & $ 3.4 \cdot 10^2$ & $1.5 $\\
\hline
$6d_{5/2}$ & $1.8 \cdot 10^2$ & $ 4.9 \cdot 10^2$ & $2.7 $\\
\hline
$5f_{5/2}$ & $1.3 \cdot 10^2$ & $ 79            $ & $0.61$\\
\hline
$5f_{7/2}$ & $65$ &  $61$ & $0.94$\\
\hline
\end{tabular}
\caption{\label{NeutrAtICRates} The $M1$ and $E2$ IC rates $\Gamma_{IC}(M1)$ and $\Gamma_{IC}(E2)$  for electrons from different electronic orbitals and their ratios.}
\end{table}

Our results show that for all electronic orbitals but $7s$, the $E2$ IC channel either dominates or is comparable to the $M1$ IC channel. For the $7p_{3/2}$ orbital the rate $\Gamma_{IC}(E2)$ becomes an order of magnitude larger than  $\Gamma_{IC}(M1)$. The $E2$ contribution can be safely neglected only for the $7s$ electron, for which selection rules in the sum over $\kappa$ in Eq.~(\ref{gammaicE2}) rule out significant contributions. We note however that all $E2$ IC decay rates in Table~\ref{NeutrAtICRates}  are in absolute value two to four orders of magnitude smaller than the $M1$ IC rate involving the $7s$ electron. This confirms that whenever the $7s$ electron is available for IC, the $M1$ IC decay channel will dominate. We recall that the IC coefficient for the $7s$ electron is  approx. $10^9$. The IC coefficients for the other orbitals are correspondingly lower, however still much larger than unity. IC from any suitable occupied orbital is therefore much more probable than the $M1$-dominated radiative decay.

\subsection{IC in Th ions}
Considering the 7.8 eV energy value for the nuclear transition, in Th ions IC becomes energetically forbidden from the electronic ground state and may only occur from electronic excited states where the $7s$ orbital is not occupied. This scenario has been studied recently in  Ref.~\cite{Bilous_PRA_2017}.  IC from excited states  for $\isotope[229]{Th}^+$ and $\isotope[229]{Th}^{2+}$ was proposed as means for characterization of the nuclear isomeric state. For the characterization scheme, IC from the initially excited state $5f6d^2$ at 30223~$\mathrm{cm}^{\mathrm{-1}}$ was considered. 
We have updated the conversion rates of the $6d$- or the $5f$-electron of the $5f6d^2$ initial state using the newly available reduced transition probability values.  The results are shown in Table~\ref{5f6d2}. The first two columns correspond to the outer electronic configuration of the $\isotope[229]{Th}^{2+}$  ion  after the IC event. Depending on the considered final state, the IC process is energetically possible only for values of the $\isotope[229]{Th}$ isomer energy $E_\mathrm{m}$ exceeding the value $E_\mathrm{m}^{\mathrm{min}}$ shown in the third column. The last three columns depict the dependence of the IC rate on $E_\mathrm{m}$. In all the cases the rate of the $E2$ channel is either comparable to the $M1$ channel or exceeding it by up to a factor three. We conclude that IC involving conversion electrons from the $6d$ and $5f$ orbitals is dominated by the $E2$ transition. For these cases, it is not justified to neglect the $E2$ component in IC calculations. 

\begin{table}[ht!]
\centering
\begin{tabular}{|c|c||c||c|c|c|}
\hline
\multicolumn{2}{|c||}{Final state} &
\multirow{2}{*}{$E_\mathrm{m}^{\mathrm{min}}$~(eV)} &
\multirow{2}{*}{$E_\mathrm{m}$~(eV)} &
\multicolumn{2}{c|}{Rate ($\mathrm{s}^{-1}$)} \\
\cline{1-2}
\cline{5-6}
Config. & $E$ ($\mathrm{cm}^{-1}$) & &  & M1 &  E2 \\
\hline
\multirow{11}{*}{$6d^2$} &
      \multirow{6}{*}{6538}   
             &
              \multirow{6}{*}{9.2}
 &9.2& 41& 38\\
&& &9.5& 41& 38\\
&& &10.0& 41& 37\\
&& &10.5& 40& 36\\
&& &11.0& 40& 35\\
&& &11.5& 39& 35\\
      \cline{2-6}
&   \multirow{5}{*}{10543}   
             &
              \multirow{5}{*}{9.7}
 &9.7& 38& 40\\
&& &10.0& 37& 39\\
&& &10.5& 37& 38\\
&& &11.0& 36& 37\\
&& &11.5& 36& 37\\
\cline{1-6}
 \multirow{19}{*}{$5f6d$}
      & \multirow{6}{*}{4490}
              & \multirow{6}{*}{9.0}
 &9.0& 37& 96\\
&& &9.5& 38& 101\\
&& &10.0& 40& 106\\
&& &10.5& 41& 112\\
&& &11.0& 43& 116\\
&& &11.5& 44& 121\\
      \cline{2-6}
&           \multirow{5}{*}{8437}   
             &
              \multirow{5}{*}{9.4}
 &9.4& 225& 405\\
&& &10.0& 224& 407\\
&& &10.5& 223& 408\\
&& &11.0& 223& 409\\
&& &11.5& 222& 410\\
      \cline{2-6}
&        \multirow{5}{*}{11277}   
             &
              \multirow{5}{*}{9.8}
 &9.8& 35& 121\\
&& &10.0& 36& 122\\
&& &10.5& 38& 126\\
&& &11.0& 40& 130\\
&& &11.5& 42& 133\\
      \cline{2-6}
&         \multirow{3}{*}{19010}   
             &
              \multirow{3}{*}{10.8}
 &10.8& 54& 159\\
&& &11.0& 54& 161\\
&& &11.5& 56& 164\\
\hline
\end{tabular}
\caption{Internal conversion rates for the state $5f6d^2$ at 30223 $\mathrm{cm}^{\mathrm{-1}}$  in $~^{229\mathrm{m}}\mathrm{Th}^{+}$. The total angular momentum of this state is $J=15/2$. \label{5f6d2}}
\end{table}

\subsection{EB in Th ions}
We present the EB enhancement factors $\beta(M1)$ and $\beta(E2)$ calculated according to Eqs.~(\ref{betas})--(\ref{betas2}) for different initial and final electronic states in a monovalent ion $\isotope{Th}^{3+}$ in Table~\ref{EBRates}. As described in more detailed above, $\beta(M1)$ is obtained with the combined DHF and RPA method, whereas $\beta(E2)$ is not sensitive to RPA corrections and is calculated directly with the DHF method. In order to directly compare our results with available theoretical values in Ref.~\cite{PorsevFlambaum_Brige3+_PRA_2010}, we have considered here the same nuclear transition energy $E_\mathrm{m}=7.6$~eV as assumed there. The agreement is satisfactory except for the value of $\beta(M1)$ for EB from the initial state  $5f_{5/2}$ to the final state $7s_{1/2}$, where the values differ by a factor 2.5. We consider this issue not critical for our purposes especially taking into account the significant disagreement in RPA calculations for $M1$ matrix elements present in literature~\cite{Gossel_PRA_2013,JohnsonSafronova_PRA_2017}.  The small deviations of the $E2$ rates can be attributed to the RPA corrections included in Ref.~\cite{PorsevFlambaum_Brige3+_PRA_2010} but missing in the present calculations.

\begin{table}[h!]
\centering
\begin{tabular}{|c|c||c|c|c||c|c|}
\hline
\multirow{2}{*}{Init.} & \multirow{2}{*}{Fin.} & \multicolumn{3}{|c||}{Present work}  & \multicolumn{2}{|c|}{Based on~\cite{PorsevFlambaum_Brige3+_PRA_2010}} \\
\cline{3-7}
 & & $\beta(M1)$ & $\beta(E2)$ & $\rho$ & $\beta(M1)$ & $\beta(E2)$ \\
\hline
$5f_{5/2}$ & $7s_{1/2}$ & 0.082 & $3.4 \cdot 10^7$ & 0.29 & 0.032 & $5 \cdot 10^7$ \\
\hline
$7s_{1/2}$ & $7p_{1/2}$ & 23 & $6.6 \cdot 10^8$ & 0.020& 19 & $7 \cdot 10^8$  \\
\hline
$7s_{1/2}$ & $7p_{3/2}$ & 5.2 & $9.3 \cdot 10^7$ & 0.012 & 4.4 & $1 \cdot 10^8$ \\
\hline
$6d_{3/2}$ & $7p_{1/2}$ & $0.0022$ & $1.2 \cdot 10^7$ & 3.8&-&- \\
\hline
$6d_{5/2}$ & $7p_{1/2}$ & 0.013 & $3.6 \cdot 10^7$ & 1.9&-&- \\
\hline
$6d_{5/2}$ & $7p_{3/2}$ & $2.4\cdot 10^{-5}$ & $2.4\cdot 10^5$ & 6.9 &-&-\\
\hline
\end{tabular}
\caption{\label{EBRates} Enhancement factors $\beta(M1)$ and $\beta(E2)$ and the ratio $\rho=\Gamma_{EB}(E2)/\Gamma_{EB}(M1)$ for $M1$ and $E2$ EB channels between different initial and final electronic states  in a monovalent  $\isotope{Th}^{3+}$ ion. For comparison we also present the $\beta(M1)$ and $\beta(E2)$ results based on DHF+RPA calculations from Ref.~\cite{PorsevFlambaum_Brige3+_PRA_2010} where available. The nuclear isomeric state energy is assumed to be $E_\mathrm{m}=7.6$~eV to match the value used in Ref.~\cite{PorsevFlambaum_Brige3+_PRA_2010}. }
\end{table}

Our results in Table~\ref{EBRates} show that $\Gamma_{EB}(E2)$ is smaller than  $\Gamma_{EB}(M1)$ only for the transitions involving the $7s_{1/2}$ orbital. In particular, $\Gamma_{EB}(E2)$ is two 
 orders of magnitude smaller than $\Gamma_{EB}(M1)$ only for the EB scheme between the electronic states $7s_{1/2}$ and $7p_{1/2,3/2}$, where  $\beta(M1)$ is largest. This can be explained by the presence of a strong $M1$ transition between the initial state $7s_{1/2}$ and the intermediate state $8s_{1/2}$. At the same time the $E2$ transition between these states is forbidden by selection rules.
However, for all other cases  the $E2$ contribution is comparable (for instance a third of the $M1$ contribution for EB between the electronic states $5f_{5/2}$ and $7s_{1/2}$) or even dominating with respect to the $M1$ channel. This supports our conclusion that the $E2$ channel should not be disregarded by default but should be considered in calculations involving coupling to the atomic shell.

\section{Conclusions}
The $E2$ multipole mixing in the isomer decay has been so far widely disregarded in the literature. Our results  based on the most recent theoretical predictions for the reduced nuclear transition probabilities $B(M1)$ and $B(E2)$  have shown that this is in many cases not justified.  The surprisingly  large $E2$ contribution to the nuclear decay channels is related to the fact that compared to previously available values, the predicted $B(M1)$ is lower while $B(E2)$ has increased. From the point of view of nuclear structure, the $E2$ component stems mainly from the strong collective QO motion of the nuclear core, with an indirect contribution to the s.p. degrees of freedom via the particle-core coupling. However, decisive for the nuclear decay mechanisms are the selection rules for the given multipolarity and the behaviour of the electronic orbital wave functions involved in the coupling to the nuclear transition.

The $M1$ decay channel dominates in the radiative decay by approx. ten orders of magnitude. This changes for the processes coupling the nuclear transition to the electronic shell, which have a more involved dependence on the multipolarity of the nuclear transition. With a  smaller factor (three orders of magnitude), $M1$ still dominates over $E2$  for IC of the $7s$ electron and EB between the electronic states $7s$ and $7p$. For IC from the other outer orbitals $7p$, $6d$, $5f$ or for EB involving other transitions than $7s\rightarrow 7p$, the $E2$ contribution is either of the same order of magnitude or even dominant compared to the $M1$ decay channel. 

Allegedly when comparing the absolute values of the IC or EB rates, the largest values are the ones for the IC $M1$ decay channel of the $7s$ electron and the EB between the electronic states $7s$ and $7p$. These will be the dominating processes whenever the electronic structure is adequate and allows them to occur. However, specific scenarios involving more exotic configurations such as IC from excited electronic states or particular EB schemes~\cite{Bilous_PRA_2017,Bilous_NJP_2018}---as followed for instance at present at the Physikalisch-Technische Bundesanstalt in Braunschweig, Germany, for exciting the $\isotope[229]{Th}$ isomer---may require consideration of the $E2$ multipole mixing contribution, which will enhance the rates of the respective processes.
Apart of the increased rate, the different selection rules applying for $E2$ may also lead to observable effects such as different angular distribution of the emitted electrons.

\begin{acknowledgments}
This work was supported by the DFG and by the BNSF under Contract DFNI-E02/6. PB and AP gratefully acknowledge funding by the nuClock EU FET-Open project 664732. 
\end{acknowledgments}

\bibliography{refs}

\end{document}